\documentclass[%
 reprint,
 amsmath,amssymb,
 aps,
]{revtex4-1}

\usepackage{graphicx}
\usepackage{dcolumn}
\usepackage{bm}

\begin{document}

\preprint{APS/123-QED}

\title{Longest-path attacks on complex networks}

\author{Cunlai Pu}
\email{pucunlai@njust.edu.cn}
\author{Wei Cui}%
\affiliation{School of Computer Science and Engineering, Nanjing University of Science and Technology, Nanjing 210094,  China
}




\date{\today}

\begin{abstract}
We investigate the longest-path attacks on complex networks. Specifically, we remove approximately the  longest simple path from a network iteratively until there are no paths left in the network. We propose two algorithms, the random augmenting approach (RPA) and the Hamilton-path based approach (HPA), for finding the approximately longest simple path in a network. Results demonstrate that steps of longest-path attacks increase with network density linearly for random networks, while exponentially increasing for scale-free networks. The more homogeneous the degree distribution is, the more fragile the network, which is totally different from the previous results of node or edge attacks. HPA is generally more efficient than RPA in the longest-path attacks of complex networks. These findings  further help us understand the vulnerability of complex systems, better protect complex systems, and design more tolerant complex systems.
\end{abstract}

\pacs{89.75.Hc}
\maketitle


\section{Introduction}
Vulnerability is one of the fundamental properties in many nature and man-made complex systems\cite{Newman09,Dor08,Barrat,albert02}. For instance, genetic defects induce cell lesions\cite{Robbins}, router failures interrupt the Internet\cite{Labovitz}, incidents or terrorist attacks cause collapse of public transport systems\cite{Elias}, and malfunction of a power station results in large-scale blackouts\cite{Kenett}. The vulnerability  of complex systems is determined by the underlying networks in which nodes represent individuals in complex systems and edges represent the interactions of individuals. In  past decades, many breakthroughs have been made on understanding individual components'  disturbance's effects on the overall function of complex networks. Albert \textit{et al}\cite{Albert00} found that the Internet and WWW (World Wide Web) are very robust against random failures, but quite fragile in  intended attacks. This robust yet fragile property was confirmed in much larger maps of the WWW as well as many other scale-free networks\cite{Newman09}, and percolation processes on model networks were introduced to explain this property\cite{Callaway,Cohen02}. Since then many researchers studied the robustness of model networks and real-world networks subject to node attacks and edge attacks\cite{Newman09,Dor08}, in which nodes or edges are removed in order of degree\cite{Cohen01,Crucitti041} or other centrality measures such as  betweenness\cite{Holme02,Iyer}, eigenvector\cite{Allesina09}, PageRank\cite{Qin}, \textit{ etc}. A few localized failures or attacks may cause cascading failures and  lead to breakdown of the whole system which has been observed in Internet\cite{Motter02,Crucitti04}, power grids\cite{Kinney05, Wang08}, financial systems\cite{Haldane}, \textit{etc}. Recently,  researchers studied robustness of temporal or time-varying networks by extending the measures of centrality and robustness with temporal properties\cite{Holme13,Scellato}. Furthermore,  current attention focuses on the percolation processes on multiplex or interdependent networks which  better model catastrophic events in power grids, transport systems and many other interdependent systems\cite{Kenett,Buldyrev,Bashan}. 
 However, attacks are  not merely limited to node or edge attacks that  have been widely studied in the literature. For instance, hurricanes always have a large-scale effect on the public transport networks, terrorists usually prefer much larger scale  attacks, and drugs always affect many targets. Therefore, we need to  understand the effect of attacks on larger parts of complex networks other than nodes and edges, like path attacks. A simple path composed of nodes and edges is a common subpart of a network. In this context, ``path"  always means ``simple path" which indicates that a node appears at most once in a path. This restriction is consistent with the attacks problem. In a network, a straightforward measure of paths is path length. Therefore, a natural question is how removals of the longest paths affect the function of a network. Finding the longest paths in general graphs is a well-known NP-hard problem in the literature\cite{Karger}. Many approximation algorithms are proposed to approximate the longest paths including color-coding method\cite{Alon}, divide and conquer approach\cite{John}, algebraic approach\cite{Koutis}, \textit{etc}.
 In this letter, we propose two algorithms, RPA and HPA,  to find the approximately longest paths during the  attack processes. The efficiency of these two algorithms in the attacks is determined by the decay rates of the largest components in networks. The robustness of model networks and real-world networks to longest-path attacks are reflected by the  steps of the iterative attacks. 
\section{Model of longest-path attacks}
In a large network, there is usually a  huge number of paths which is much larger than the number of nodes and edges. In some sense, paths can be thought of as  the combinations of nodes and edges. When attacking a network, we prefer to remove the critical paths so that the removal of these paths significantly degenerate the function of the network. There are plenty of centrality measures in literature\cite{Newman09}, but most of them are for nodes and edges, few are connected with paths. Path length, a natural measure for a path, is defined as the number of edges in a path. In our path-attack process, each time we remove the longest path from the network based on specified algorithms. Note that when attacking a path, we just remove all its edges from the network, but keep all its nodes in the network. The removal of longest path continues until there are no paths in the networks. In fact there are no edges in the final network, since edges are paths of length 1, and they would be removed from the network if there were any left. 
\section{Measures of longest-path attacks}
The size of the largest component in a network reflects the communication capability of a network. Therefore we compute the size of largest component $S$ during each step of the longest-path attacks, which indicates the efficiency of an approximating algorithm. The larger the decay rate of the largest component, the more efficient the approximating algorithm. Also, we count the total step of the longest-path attacks $T$, which reflects the tolerance of a network subject to longest-path attacks, and more steps means a more robust  network against longest-path attacks.
\section{ Algorithm for approximating the longest path}
In each iteration of the attacks process, we need to find the longest path in the corresponding network, which is a famous NP-hard problem. Here we present two algorithms,  random path augmenting approach (RPA) and Hamilton-path based augmenting approach (HPA), to find the approximate longest path in a network. 
\subsection{ Random path augmenting approach}
RPA is a very simple algorithm for constructing a path in a network. First, RPA randomly selects a node from the network as the root node. Second, RPA chooses a neighboring node of the root node as the second node of the path, and then chooses a neighboring node of the second node as the third node of the path, and so on. The path augmenting process continues until it cannot find a new appropriate node. The main procedure of RPA is as follows:
\begin{enumerate}
\renewcommand{\labelenumi}{(\theenumi)}
\item Maintain a path $P$ which is a node sequence (initially null), and node set $V_0$ (initially null).
\item Randomly select a node $ r $ as the root node.
\item Append $r$ to the end of $P$, and add $r$ to $V_0$.
\item Find a neighboring node of $r$ named $q$, such that $q\notin V_0$.
\item IF $q$ exists, take $q$ as $r$, return to (3). Otherwise, the algorithm stops.
\end{enumerate}
\begin{figure}
 \centering
\includegraphics[width=3in,height=3in]{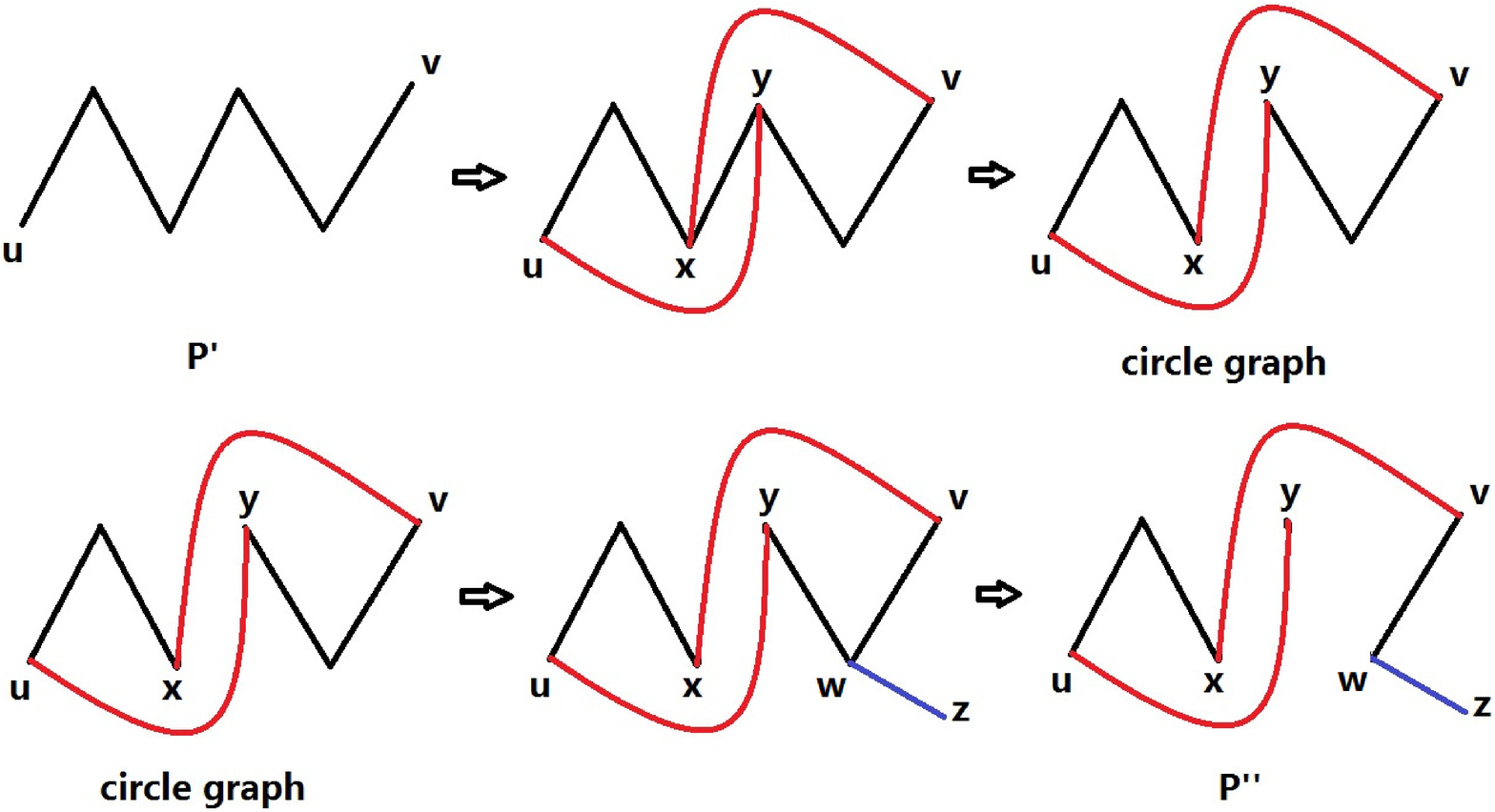}
\caption{An example for illustrating the idea of HPA. }
\end{figure}
\begin{figure}
\centering
\includegraphics[width=3.4in,height=1.7in]{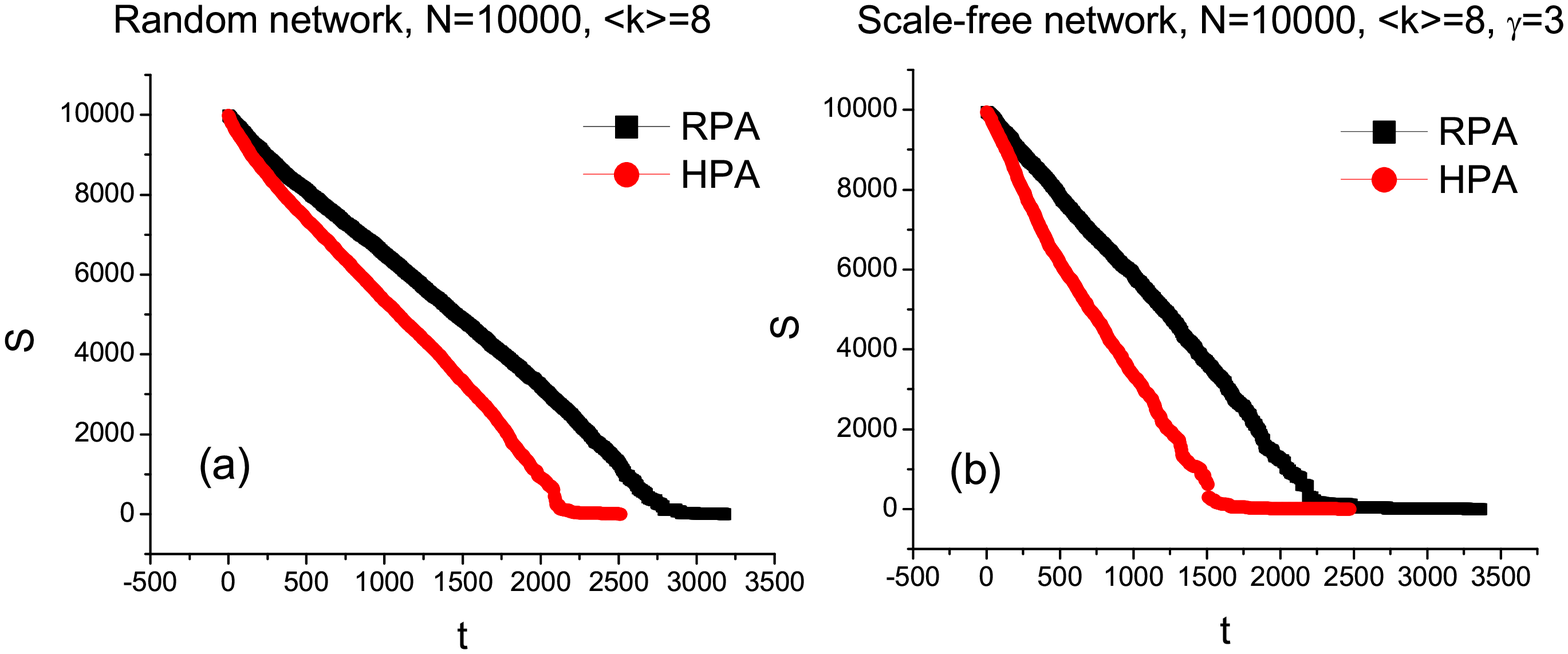}
\caption{Size of the largest component $S$ vs. time step of the longest-path attack $t$ for (a) the random network and (b) the scale-free network. }
\end{figure}
\subsection{Hamilton-path based augmenting approach} 
HPA employs the idea of approximating the Hamilton path in a network\cite{Itai}. Initially, HPA randomly selects a node as the root node, and augments a path as long as possible in both directions of the root node. In each iteration, HPA first generates a circle graph based on the current augmenting path. Then HPA makes the circle graph a new augmenting path by introducing an arbitrary node connected to the circle graph, and starts a new augmenting process.
The procedure of HPA is as follows:
\begin{enumerate}
\renewcommand{\labelenumi}{(\theenumi)}
\item Maintain a path $P$ that is a node sequence (initially null).
\item Randomly select a node $ r$ as the root node, and let $P= \langle r \rangle$. 
\item Augment a path in both directions from the current path $P$ as far as possible, and obtain a new path $P'$. Assuming $P=\langle u0,\cdots,v0\rangle$, then $P'=\langle u,\cdots,u0,\cdots,v0,\cdots,v \rangle$. Note that a node is allowed to appear once in a path.
\item Find an edge $ \langle x, y \rangle$ in $P'$ such that edge $\langle u, y\rangle$ and edge $\langle x, v \rangle$ exist. If edge $\langle x, y \rangle$ is found, then delete it, and add edge $\langle u, y \rangle$ and edge $\langle x, v \rangle$ to $P'$, which makes $P'$  a circle graph, shown in the fig. 1. If edge $\langle x, y \rangle$ is not found, the algorithm ends, and $P'$ is the resulting path. Note that if the network satisfies Dirac constraints and $P'$ is not a Hamilton path, then $\langle x, y \rangle$ exists (see theorem 1 in Appendix A).
\item Find a node $z$,  such that $z$ is not included in the circle graph, but $z$ connects to an arbitrary node $w$ in the circle. If $z$ is found, then delete an arbitrary edge incident to $w$ from the circle graph, and add edge $\langle z, w \rangle$ to the circle graph. As a result, the circle graph becomes a longer path $P''$, as shown in fig. 1. If $z$ is not found, the algorithm ends, and $P' $ is the resulting path. Note that if the network satisfies the Dirac constraints and $P'$ is not a Hamilton path, then $z$ exits (see theorem 2 in Appendix B).
\item Take $P''$ as $P$, and return to (3).
\end{enumerate}
The circle graph and the augment path, generated in the intermediate stage of HPA, become larger  with the increasing  iteration times. We  infer that if the network is dense enough (satisfying the Dirac constraints), then the resulting path explored by HPA is a Hamilton path, which is  exactly the longest path in the network. 
\subsection{Space and time complexity of the two algorithms}
Assume there are $n$ nodes and $m$ edges in a network. For one iteration, RPA runs in $O(n)$ space constantly and runs in $O(n)$ time in the worst case, while HPA runs in $O(n)$ space and $O(n^2)$ time in the worst case. For many iterations, RPA only runs in $O(m)$ time, while HPA only runs in $O(\lfloor m/n\rfloor \ast n^2)$ time,  which are irrelevant with the specific iteration times. More detailed analysis of computing complexity for RPA and HPA is presented in Appendix C.
\section{Simulation results}
In this part, we show the results of longest-path attacks for model networks and real-world networks. The Erd\"{o}s-R\'{e}nyi (ER) model\cite{Erdos} and the static model\cite{Goh01} are employed to generate the random networks and scale-free networks respectively. The statistics of the real-world networks are shown in table. 1.
\subsection{Model networks}
We first investigate decrease of size of the largest component $S$ during the iterative longest-path attacks. In fig. 2, we clearly see that, $S$ decreases with time step $t$ almost linearly, and fast in the first stage,  and slowly in the final stage. The behavior of $ S$ is similar in both the random network and the scale-free network. The efficiency of RPA and HPA is different which is inferred from the corresponding decay rates of $S$.  HPA is more efficient than RPA since HPA has a much smaller $S$ for a specific $t$,  which is found in both the random network and the scale-free network.
\begin{figure}
 \centering
\includegraphics[width=3.4in,height=1.7in]{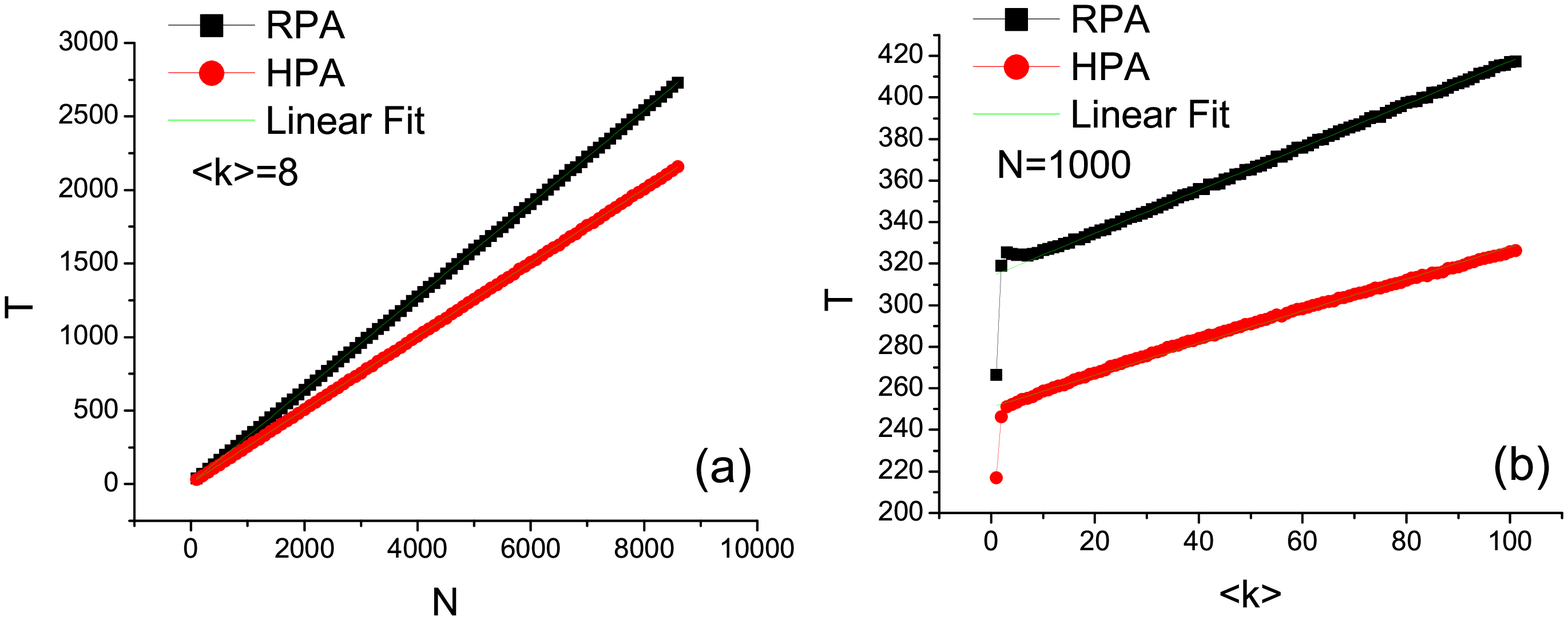}
\caption{(a) Total attack steps $T$ vs. network size $N$, and (b) total attack steps $T$ vs. average degree $\langle k \rangle$ for random networks. The results are the average of $10^{4}$ independent runs.}
\end{figure}

Then we investigate the total attack steps $T$ of model networks for the two  algorithms, shown in fig. 3 and fig. 4. $T$ increases linearly with network's size $N$ for the two algorithms on both random networks (fig. 3(a)) and scale-free networks (fig. 4(a) and fig. 4(b)). For a given $N$, $T$ of HPA is always  smaller than that of RPA.
 $T$ increases with average degree $\langle k \rangle$ linearly for random networks (fig. 3(b)), while  exponentially for scale-free networks (fig. 4(c) and fig. 4(d)).  However, when the power-law parameter $\gamma=7$, $T$ increases linearly as shown in fig. 4(c) and fig. 4(d).  This is because the heterogeneity of the degree distribution of scale-free networks decrease with the power-law parameter $\gamma$\cite{Goh01}. Therefore, when $\gamma$ is large enough, the scale-free networks are actually random networks.   The increase of $T$ with $\langle k \rangle$ indicates that the denser the network, the more robust the network is to longest path attacks. Also for specific $\langle k \rangle$, $T$ of HPA is always smaller than that of RPA.
From fig. 4(e) and fig. 4(f), we see that $T$ decreases with the power-law parameter $\gamma$ of scale-free networks. This means that the more homogeneous the degree distribution, the more fragile the network is  to longest-path attacks, which is completely opposite to the previous results of node or edge attacks\cite{albert02,Albert00}. Furthermore, we obtain from fig. 3 and fig. 4 that the  networks are generally more robust to RPA  than to HPA, which is inferred from the behavior of  $T$, and holds for both random networks and scale-free networks. 
\begin{figure*}
 \centering
\includegraphics[width=3.5in,height=3in]{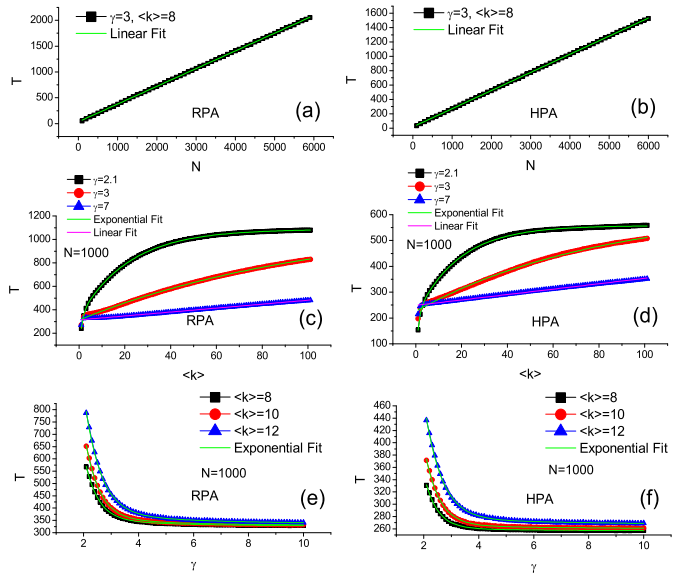}
\caption{Total attack steps $T$ vs. network size $N$ ( (a) (b) ), total attack steps $T$ vs. average degree $\langle k \rangle$ ( (c) (d) ), total attack steps $T$ vs. power-law parameter $\gamma$ ( (e) (f) ) for scale-free networks. The results are the average of $10^{4}$ independent runs.}
\end{figure*}
\begin{figure}
 \centering
\includegraphics[width=2.65in,height=3in]{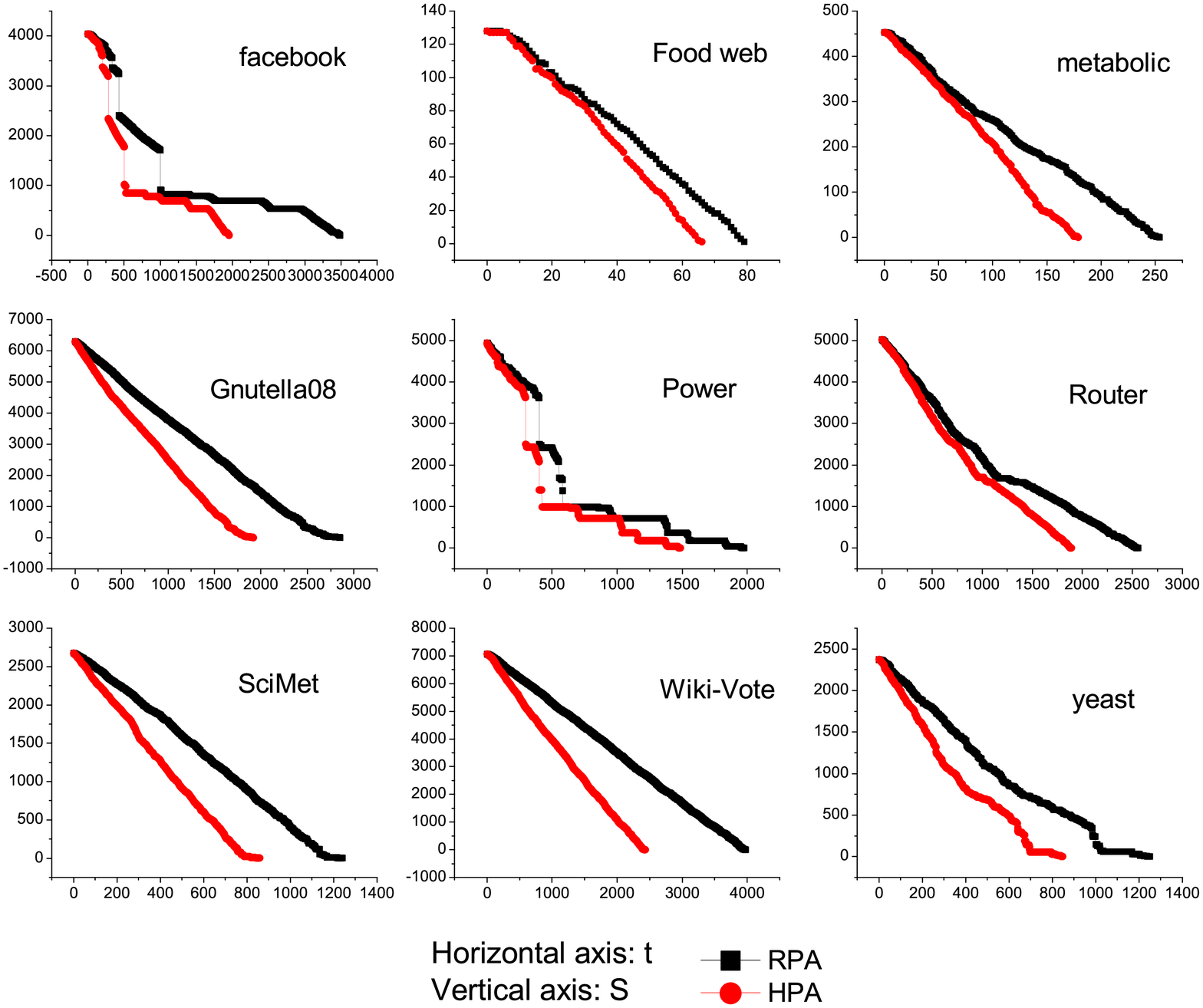}
\caption{Size of the largest component $S$ vs. time step of the longest-path attack $t$ for real-world networks.}
\end{figure}
\subsection{Real-world networks}
We also demonstrate the behavior of the largest component $S$ in the longest-path attacks and the total steps of attacks $T$ for real-world networks (table 1). In fig. 5, generally the size of the largest component   $S$ decreases  with time step $t$ linearly, which is similar in model networks. However,  the fluctuations  in the curves of some real-world networks are clear.  HPA is   more efficient than RPA for real-world networks, since HPA always has a much smaller $S$ for a specified $ t$ shown in fig. 5. The robustness of  real-world networks to longest-paths attacks is reflected by the corresponding total steps of attacks $T$, shown in table 1. We obtain that  real-world networks are more fragile to HPA than to RPA, since HPA has a much smaller $T$ than RPA. 
\begin{table}
\caption{Attack steps $T$ for real-world networks.}
\begin{center}
\begin{tabular}{lcccr}
\hline
\hline
NAME&NODES&EDGES&RPA&HPA\\
facebook&4039&88234&3486&1953\\
Food web&128&2106&79&66\\
Kohonen&4470&12731&1874&1214\\
metabolic&453&4596&254&179\\
Gnutella08&6301&20777&2853&1920\\
Power&4941&6594&1976&1487\\
Router&5022&6258&2555&1890\\
Wiki-Vote&7115&103689&3976&2435\\
Yeast&2375&11693&1250&848\\
SciMet&3084&10416&1240&858\\
USAir&1532&2126&233&147\\
polbooks&105&441&63&43\\
karate&34&78&22&14\\
Jazz&198&5484&112&82\\
C. elegans&297&2359&27&22\\
adjnoun&112&425&58&40\\
\hline
\end{tabular}		
\end{center}
\end{table}
\section{Conclusions and discussions}
In summary, we investigate the robustness of complex networks, including model networks and real-world networks, which are subject to iterative attacks on the longest paths. Two algorithms are proposed to approximate the longest paths during the attacks. We compare the efficiency of the two algorithms by the decay rate of the largest components, and determine the robustness of the networks by the total attack steps  in the longest-path attacks process. Specifically, we obtain that, total attack steps increase exponentially with network density for scale-free networks, while  remaining  linear for random networks. Total attack steps increase linearly with network size for both random networks and scale-free networks. For scale-free networks, the more heterogeneous the degree distribution, the more robust the networks are against longest-path attacks, which are totally different from the previous results of node or edge attacks. 
Note that finding the exact longest simple path in a general network is NP-hard. We only approximate the longest paths using two approximating algorithms. There should be more efficient approximating algorithms emerging in the future. We only investigate effects of the longest paths on the robustness of complex networks. However, our work may shed some light on the influence of longest path on the spread\cite{Zhou06, Yang13}, synchronization\cite{Yan07}, control\cite{Yan12,Yuan}, and other network dynamics.
\acknowledgments
The authors  thank   Jian Yang,  Andrew Michaelson and Siyuan Li  for
their valuable  suggestions.
This work was  supported by the Natural Science Foundation of China (Grant No. 61304154), the Specialized Research Fund for the Doctoral Program of Higher Education of China  (Grant No. 20133219120032), and the Postdoctoral Science Foundation of China (Grant No. 2013M541673).
\appendix
\section{Theorem 1}
\textbf{Theorem 1}: For any node $s$ in the network, if $d(s) \geq \lceil |V|/2 \rceil$ (Dirac constraints), $d(s)$ is the degree of $s$, $V$ is the node set of the network, then the edge $\langle x, y \rangle$ in step (4) of HPA exists.

\textbf{Proof}: According to the constraints, we have $ d(u) \geq \lceil |V|/2 \rceil$, and $d(v) \geq \lceil |V|/2 \rceil$. Since $u$ and $v$ are the two end nodes of $P'$, which is the intermediate maximum augmenting path, all the neighbors of $u$ and $v$ should be on $P'$. Based on the inclusion-exclusion principle, we can easily obtain that edge $\langle x,y \rangle$ exists. The explanation is as follows:
\begin{enumerate}
\renewcommand{\labelenumi}{(\theenumi)}
\item If we take the node positions  in $P'$  as vacancies, then there are at most $|V|$ vacancies denoted by $a_1,a_2,\dots,a_{|V|}$. Once an arbitrary neighbor node of $u$ takes one of the vacancies $a_i$, then $a_{i-1}$  can not be occupied by  any neighbor node of $v$. (Otherwise, nodes in  $a_{i-1}$ and $a_i$ form the node pair  $\langle x, y \rangle$ we need.)
\item Since $d(u) \geq \lceil |V|/2 \rceil$,  the neighbor nodes of $u$  take at least $\lceil|V|/2\rceil$ vacancies. According to (1), there are at least $\lceil|V|/2\rceil$ vacancies which cannot be occupied by $v$'s neighbor nodes. On the other hand, the maximal number of vacancies for $v$'s neighbor nodes is $|V|-1$ in theory. Therefore, the number of remaining vacancies for the neighbor nodes of $v$ are smaller than  $\lceil|V|/2\rceil$. However, $v$'s neighbor nodes need no less than  $\lceil|V|/2\rceil$ vacancies, since  $d(v) \geq \lceil |V|/2 \rceil$. This means that $v$'s neighbor nodes have to occupy at least one of the vacancies such as $a_j$, which is  forbidden for  $v$'s neighbor nodes. Then nodes in $a_j$ and $a_{j+1}$ form the node pair $\langle x, y \rangle$ we need.
\end{enumerate}
\section{Theorem 2}
\textbf{Theorem 2}: For any node $s$ in the network,  if $d(s) \geq \lceil |V|/2 \rceil$ (Dirac constraints), $d(s)$ is the degree of $s$, $V$ is the node set of the network,  then the node $z$ in step (5) of HPA exists.

\textbf{Proof}:  Assume there is no such node $z$ that connects the circle, then the circle is an independent subgraph of $G$ (the whole network), which means the circle is separated from $G$-circle.
Suppose  there are $n_0$ nodes in the circle. For any node $i$ in the circle, $d(i)\leq n_0-1$. For any node $j$ in $G$-circle, $d(j)\leq (|V|-n_0)-1$.    
Then $d(i)+d(j)\leq|V|-2$, which means $d(i)$ or $d(j)$ is less than $\lceil |V|/2 \rceil$.This contradicts the  constraints.
\section{Analysis of space and time complexity}
We assume that an undirected and unweighted graph is denoted by $G\langle V, E\rangle$, $V$ is the node set, and $E$ is the edge set of the graph. Suppose $ n=|V|$, $m=|E|$, and $p=[m/n]$  indicates the density of the graph. Obviously $0\leq p \leq n$.

1) \textbf{Random path augmenting approach}

RPA is a simple algorithm for  augmenting a path. Obviously it has $O(n) $ space complexity. For one iteration, RPA runs in $O(n)$ time,  in the worst case. Also, the number of iterations of the longest-path attacks is at most $O(n)$. Let us assume the number of iterations is $T$, the length of the resulting path in the $i$-th iteration is $A_i$ (Time complexity is $O(A_i)$). Since, $m = \sum_{i=1}^{T}A_i$, we obtain the total time complexity for RPA is $O(A_1)+O(A_2)+...+O(A_T)=O(m)$. 

2) \textbf{Hamilton-path based augmenting approach} 

  HPA runs in $O(n)$ space in the worst case. Time complexity for one iteration is $O(n^2)$.  The number of iterations is at most $O(m)$. Assume the number of iterations is $T$, the length of the explored path in the $i$-th iteration is $A_i$. Then the time complexity for the $i$-th iteration varies from $O(A_i)$ to $O(A_i^2)$. Since $m = \sum_{i=1}^{T}A_i$, we obtain that the overall time complexity $TC$ that satisfies the following constraints of quadratic programming:
\begin{enumerate}
\renewcommand{\labelenumi}{(\theenumi)}
\item $0<T \leq m$;
\item $0<A_i \leq n, i=1,...,T$;
\item $m = \sum_{i=1}^{T}A_i$;
\item $TC=O(\sum_{i=1}^{T}A_i^2 )$.
\end{enumerate}
When the distribution of $A_i$ is extremely uneven, HPA has the worst time complexity, which is calculated as follows:
\begin{eqnarray}
TC \nonumber
&=& O( \sum_{i=1}^{T}A_i^2 ) \\ \nonumber 
　　&\leq& O( [m/n]\ast n^2 + (T-[m/n]) ) \\ \nonumber 
& = &O( [m/n]\ast n^2)\\ 
& = &O( p\ast n^2 ) 
\end{eqnarray}
According to eq. C1, we obtain the following  conclusions:
\begin{enumerate}
\renewcommand{\labelenumi}{(\theenumi)}
\item The worst time complexity is irrelevant with the number of iterations of attacks, and it only depends on network scale and network density.
\item The worst time complexity for one iteration is $O(n^2)$. For many iterations, the total time complexity grows at most by an order of $O(p)$ magnitude, obviously $O(p)\leq O(n)\leq O(m)$.
\end{enumerate}

\end{document}